\documentclass[article, 10pt]{IEEEtran}

\usepackage{graphicx}
\usepackage{color}
\usepackage{placeins}
\usepackage{float}
\usepackage{hyperref}
\usepackage{tabularx,colortbl}
\usepackage{amsmath}
\usepackage{psfrag}
\usepackage{pstool}

\begin{document}
%
\title{Scattering at Interluminal Interface}

\author{\IEEEauthorblockN{Zo\'e-Lise Deck-L\'eger and Christophe Caloz}
\IEEEauthorblockA{Polytechnique Montr\'eal, Montr\'eal, QC, Canada\\ zoe-lise.deck-leger@polymtl.ca, christophe.caloz@polymtl.ca}}

\maketitle

\begin{abstract}
Interest for spacetime structures has recently been revived, following developments in metamaterials and ultrafast optics. Such structures essentially consist of successions of space-time interfaces for which the theory is still incomplete, in particular in the regime where the interface velocity lies between the wave velocities in the two media involved. This paper addresses this ``interluminal'' regime, providing exact scattering solutions for both approaching and receding interfaces. The solutions are verified to be consistent with the transmission matrix perspective and to be continuous at the limits of the subluminal and superluminal regimes.
\end{abstract}

\section{Introduction}

Spacetime structures, which are characterized by medium parameters that vary in both space and time, possess many interesting and useful properties: they are intrinsically nonreciprocal, due to their time-reversal asymmetry, induce spatial and temporal frequency transformations, provide amplification or attenuation, and support exotic phenomena such as time reversal and anomalous deflections. Combined with recent advances in metamaterial and fabrication technologies, these properties have recently revived the  interest in such media.

Medium interfaces constitute the building blocks of spacetime composite structures. The scattering coefficients and frequency transitions occurring at such interfaces are easily found from field continuity conditions, except in the regime where the interface speed is between the wave speeds in the two involved media. No complete exact solution has ever been reported in this ``interluminal'' regime.

This regime was first identified as problematic in the study of space-time periodic structures in 1963~\cite{cassedy1963dispersion}, where it was called the ``sonic'' regime. Later, the problem of an interface approaching incident light from an electromagnetic rarer medium to an electromagnetic denser medium was solved in~\cite{ostrovskii1967correct,ostrovskii1975}. This problem was investigated using the finite-difference technique in~\cite{biancalana2007dynamics} and an ingenious space-time discretization provided a solution for acoustic waves in~\cite{shui2014one}.

This paper first revisits the solution of~\cite{ostrovskii1967correct}, and then solves the problem of a receding interface, which was not treated before. The solution is shown to be consistent with the general transfer matrix perspective presented in~\cite{biancalana2007dynamics}, and to exhibit proper continuity at the limits of the velocity range.
\section{Interluminal Problems}
The interluminal regime corresponds to the velocity range
\begin{equation}\label{eq:regime}
  v_2<|v_\text{m}|<v_1,
\end{equation}
where $v_\text{m}$ is the velocity of the interface and $v_{1,2}=c/n_{1,2}$ are the speeds of light in media~1 and~2. Here we shall consider, for simplicity, the case $n_1<n_2$, with $n_1,n_2\geq1$.

Figures 1(a) and (c) present the space-time diagrams of an approaching ($v<0$) and a receding ($v>0$) interface of interluminal velocity. A wave incident on the approaching interface generates three scattered waves, while a single scattered wave is produced by three waves incident on the receding interface. These problems are under- and over-determined, respectively, since they involve three and one unknowns, for two continuity conditions~\cite{ostrovskii1975}.
\begin{figure}[h]
\centering
\psfragfig[width=0.8\columnwidth]{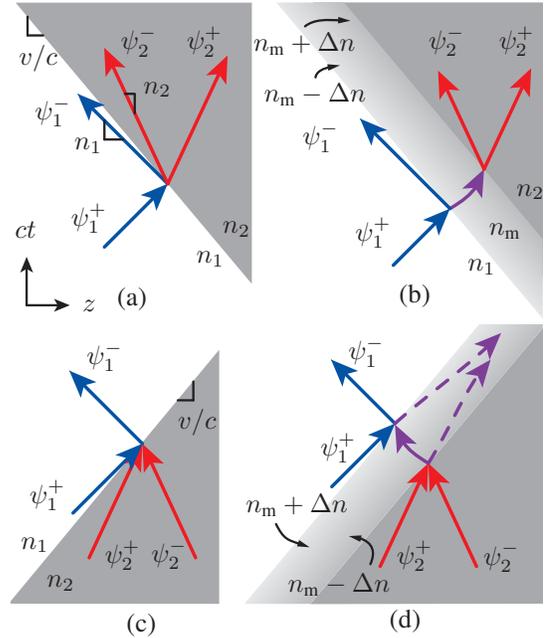}{
\psfrag{z}[c][c]{$z$}
\psfrag{t}[c][c]{$ct$}
\psfrag{v}[c][c]{$n_\text{m}-\Delta n$}
\psfrag{V}[c][c]{$n_\text{m}+\Delta n$}
\psfrag{m}[c][c]{$v/c$}
\psfrag{M}[c][c]{$v/c$}
\psfrag{n}[c][c]{$n_1$}
\psfrag{N}[c][c]{$n_2$}
\psfrag{A}[c][c]{$n_1$}
\psfrag{B}[c][c]{$n_2$}
\psfrag{C}[c][c]{$n_\text{m}$}
\psfrag{a}[c][c]{(a)}
\psfrag{b}[c][c]{(b)}
\psfrag{c}[c][c]{(c)}
\psfrag{d}[c][c]{(d)}
\psfrag{1}[c][c][1]{$\psi_1^+$}
\psfrag{2}[c][c][1]{$\psi_1^-$}
\psfrag{4}[c][c][1]{$\psi_2^+$}
\psfrag{3}[c][c][1]{$\psi_2^-$}
\psfrag{5}[c][c][1]{$\psi_3^+$}
}
\centering
\caption{Scattering from an interluminal interface. (a)~Original problem of approaching ($v<0$) interface. (b)~Auxiliary approaching proble.(c)~Original problem of receding ($v>0$) interface. (b)~Auxiliary receding problem.}
\label{fig:n12}
\end{figure}

We solve these problems using the auxiliary problems in Fig.~1(b) and Fig.~1(d). In essence, the auxiliary problems avoid the aforementioned indeterminacies by adding a gradient slab at the interfaces. This transforms the problems into successions of purely subluminal and superluminal interfaces, for which analytical solutions are known: the transmission and reflection coefficients for a subluminal ($|v_\text{m}|<v_j$) interface between media $i$ and $j$ are
\begin{subequations}\label{eq:coeff_sub}
\begin{equation}\label{eq:tauji}
\tau_{ji}=\frac{A_j^+}{A_i^+}=
\frac{2n_i}{n_i+n_j}\left(\frac{1-n_iv_\text{m}/c }{1-n_jv_\text{m}/c }\right),
\end{equation}
\begin{equation}\label{eq:gammaii}
\gamma_{ii}^{i\rightarrow j}=\frac{A_i^-}{A_i^+}=
\frac{n_i-n_j}{n_i+n_j}\left(\frac{1-n_iv_\text{m}/c }{1+n_iv_\text{m}/c }\right),
\end{equation}
\end{subequations}
while for the superluminal ($|v_\text{m}|>c/n_i$) case, they are~\cite{ostrovskii1967correct}
\begin{subequations}\label{eq:coeff_sup}
\begin{equation}\label{eq:xiji}
\xi_{ji}=\frac{A_j^+}{A_i^+}=\frac{n_i+n_j}{2n_j}\left(\frac{1-n_iv_\text{m}/c}{1-n_jv_\text{m}/c}\right),
\end{equation}
\begin{equation}\label{eq:zetaji}
\zeta_{ji}=\frac{A_j^-}{A_i^+}=-\frac{n_i-n_j}{2n_j}\left(\frac{1-n_iv_\text{m}/c}{1+n_jv_\text{m}/c}\right),
\end{equation}
\end{subequations}
where the reflective superscript $i\rightarrow j$ denotes incidence from $i$ towards $j$.

\section{Scattering from Approaching Interface}

Consider the auxiliary problem of the approaching interface, shown in Fig.~\ref{fig:n12}(b). The intermediate gradient medium has an average index $n_\text{m}=c/|v_\text{m}|$, varying from $n_\text{m}-\Delta n$ to $n_\text{m}+\Delta n$. The first interface is subluminal, since $|v_\text{m}|<c/(n_\text{m}-\Delta n)$, while the second interface is superluminal, since $|v_\text{m}|>c/(n_\text{m}+\Delta n)$. The total scattering coefficients are obtained by multiplying the coefficients of the first and second interfaces, then letting $\Delta n\rightarrow 0$, and finally inserting $v_\text{m}/c=-n_\text{m}$, which yields
\begin{subequations}\label{eq:coeff_approach_luminal_interface}
\begin{equation}
\alpha_{11}^{1\rightarrow 2}=\frac{\psi_1^-}{\psi_1^+}=\gamma_{11}^{1\rightarrow \text{m}}=-1,
\end{equation}
\begin{equation}
\quad\beta_{21}=\frac{\psi_2^-}{\psi_1^+}=\zeta_{2\text{m}}\tau_{\text{m}1}=-\frac{n_1}{n_2},
\end{equation}
\begin{equation}
\quad\kappa_{21}=\frac{\psi_2^+}{\psi_1^+}=\xi_{2\text{m}}\tau_{\text{m}1}=\frac{n_1}{n_2}.
\end{equation}
\end{subequations}
Surprisingly, these coefficients are independent of velocity.

\section{Scattering from Receding Interface}

Consider now the receding-interface auxiliary problem, shown in Fig.~1(c). There can be up to three incident waves, but only one wave can be scattered. Adding the same transition medium as previously, we  have again a succession of subluminal and superluminal problems. Using the same procedure as above, we obtain
\begin{subequations}\label{eq:coeff_recede_luminal_interface}
\begin{equation}\label{eq:coeff_recede_a}
\alpha_{11}^{1\rightarrow2}=\frac{\psi_1^-}{\psi_1^+}=\gamma_{11}^{1\rightarrow\text{m}}=-\left(\frac{1-v_\text{m}n_1}{1+v_\text{m}n_1}\right)^2,
\end{equation}
\begin{equation}\label{eq:coeff_recede_b}
\beta_{12}=\frac{\psi_1^-}{\psi_2^+}=\bar{\tau}_{1\text{m}}\zeta_{\text{m}2}=-\left(\frac{1-v_\text{m}n_2}{1+v_\text{m}n_1}\right)^2,
\end{equation}
\begin{equation}\label{eq:coeff_recede_c}
\kappa_{12}=\frac{\psi_1^-}{\psi_2^-}=\bar{\tau}_{1\text{m}}\bar{\xi}_{\text{m}2}=\left(\frac{1+v_\text{m}n_2}{1+v_\text{m}n_1}\right)^2,
\end{equation}
\end{subequations}
where $\bar{\tau}=\tau(-v_\text{m})$ and $\bar{\xi}=\xi(-v_\text{m})$. Note the two purple dashed lines never escape from the transition region, and therefore do not enter into the calculation of the coefficients.

\section{Verification}

We now check whether our results are consistent with the transfer matrix of a spatiotemporal interface~\cite{biancalana2007dynamics,Deck_arXiv_2018}, which relates the fields on both sides of the interface:
\begin{equation}\label{eq:TM}
\begin{bmatrix}
\psi_i^+\\
\psi_i^-
\end{bmatrix}=\begin{bmatrix}
\dfrac{n_j+n_i}{2n_i}\dfrac{1+n_jv_\text{m}}{1+n_iv_\text{m}}&\dfrac{n_j-n_i}{2n_i}\dfrac{1-n_jv_\text{m}}{1+n_iv_\text{m}}\\[2ex]
\dfrac{n_j-n_i}{2n_i}\dfrac{1+n_jv_\text{m}}{1-n_iv_\text{m}}&\dfrac{n_j+n_i}{2n_i}\dfrac{1-n_jv_\text{m}}{1-n_iv_\text{m}}\\[2ex]
\end{bmatrix}\begin{bmatrix}
\psi_j^+\\
\psi_j^-
\end{bmatrix}.
\end{equation}
Reading out the field relationship from Fig.~1(c) and replacing the fields in medium~1 using~\eqref{eq:TM} yields
\begin{equation}\label{eq:TM_verif}
\begin{split}
\kappa_{12}\psi_2^+&+\beta_{12}\psi_2^-=\psi_1^--\alpha_{11}^{1\rightarrow 2}\psi_1^+\\
&=\left(t_{12}^{21}-\alpha_{11}^{1\rightarrow 2}t_{12}^{11}\right)\psi_2^++\left(t_{12}^{22}-\alpha_{11}^{1\rightarrow 2}t_{12}^{12}\right)\psi_2^-,
\end{split}
\end{equation}
where $t_{12}^{nm}$ corresponds to the transfer matrix element of the $n$\textsuperscript{th} row and $m$\textsuperscript{th} column. We find our results are consistent with this transfer matrix, since inserting $\alpha_{11}^{1\rightarrow 2}$ ~\eqref{eq:coeff_recede_a} and the transfer matrix elements of~\eqref{eq:TM} into the right side of~\eqref{eq:TM_verif} retrieves the coefficients $\kappa_{12}$ \eqref{eq:coeff_recede_b} and $\beta_{12}$ \eqref{eq:coeff_recede_c}.

\begin{figure}[h]
\centering
\psfragfig*[width=1\columnwidth]{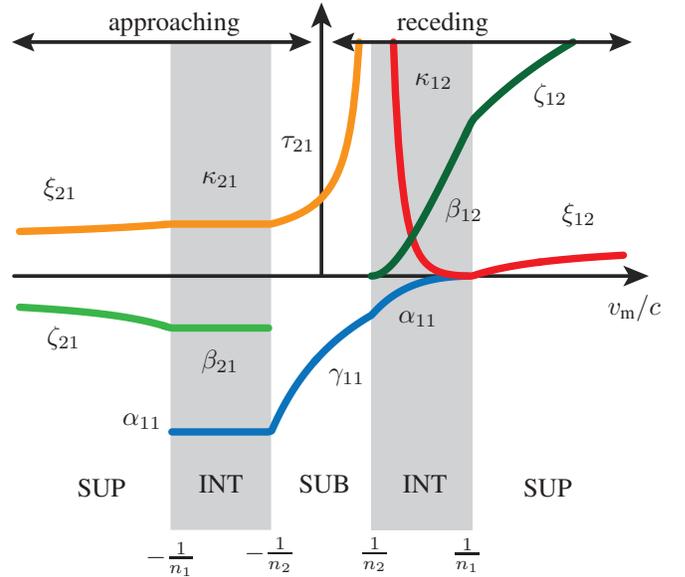}{
\psfrag{a}[c][c]{$\xi_{21}$}
\psfrag{b}[c][c]{$\kappa_{21}$}
\psfrag{c}[c][c]{$\tau_{21}$}
\psfrag{d}[c][c]{$\kappa_{12}$}
\psfrag{e}[c][c]{$\xi_{12}$}
\psfrag{A}[c][c]{$\zeta_{21}$}
\psfrag{B}[c][c]{$\beta_{21}$}
\psfrag{C}[c][c]{$\beta_{12}$}
\psfrag{D}[c][c]{$\zeta_{12}$}
\psfrag{x}[c][c]{$\alpha_{11}$}
\psfrag{y}[c][c]{$\gamma_{11}$}
\psfrag{u}[c][c]{$-\frac{1}{n_1}$}
\psfrag{U}[c][c]{$-\frac{1}{n_2}$}
\psfrag{1}[c][c]{$\frac{1}{n_1}$}
\psfrag{2}[c][c]{$\frac{1}{n_2}$}
\psfrag{v}[c][c]{$v_\text{m}/c$}
\psfrag{s}[c][c]{SUP}
\psfrag{S}[c][c]{SUB}
\psfrag{i}[c][c]{INT}
\psfrag{r}[c][c]{receding}
\psfrag{n}[c][c]{approaching}
}
\centering
\caption{Scattering coefficients [Eqs.~\eqref{eq:coeff_sub}, \eqref{eq:coeff_sup}, \eqref{eq:coeff_approach_luminal_interface} and \eqref{eq:coeff_recede_luminal_interface}] across the entire velocity range, which includes the approaching and receding subluminal (SUB), interluminal (INT) and superluminal (SUP) regimes.}
\label{fig:coefs}
\end{figure}

To further verify our results, the scattering coefficients are plotted in Fig.~\ref{fig:coefs} across the entire velocity range. It is comforting to see the parameters exhibit no discontinuity at the different regime limits.

\bibliography{interluminal_interface-arXiv.bbl}
\end{document}